\documentclass[sn-aps]{sn-jnl}% American Physical Society (APS) Reference Style
\usepackage{graphicx}%
\usepackage{multirow, subcaption, braket, hyperref}%
\usepackage{amsmath,amssymb,amsfonts}%
\usepackage{amsthm}%
\usepackage{mathrsfs}%
\usepackage[title]{appendix}%
\usepackage{xcolor}%
\usepackage{textcomp}%
\usepackage{manyfoot}%
\usepackage{booktabs}%
\usepackage{algorithm}%
\usepackage{algorithmicx}%
\usepackage{algpseudocode}%
\usepackage{listings}%
\usepackage{soul}
\begin{document}

\title[Article Title]{Mitigating imperfections in Differential Phase Shift Measurement-Device-Independent Quantum Key Distribution via Plug-and-Play architecture}


\author*[1]{\fnm{Nilesh} \sur{Sharma}}\email{ee20d060@smail.iitm.ac.in}
\author[3]{\fnm{Shashank Kumar} \sur{Ranu}}
\author[2]{\fnm{Prabha} \sur{Mandayam}}
\author[1]{\fnm{Anil} \sur{Prabhakar}}

\affil*[1]{\orgdiv{Dept. of Electrical Engg.}, \orgname{IIT Madras}, \orgaddress{\city{Chennai}, \postcode{600036}, \state{Tamil Nadu}, \country{India}}}

\affil[2]{\orgdiv{Dept. of Physics}, \orgname{IIT Madras}, \orgaddress{\city{Chennai}, \postcode{600036}, \state{Tamil Nadu}, \country{India}}}

\affil[3]{\orgdiv{Dept. of Mathematics}, \orgname{University of York}, \orgaddress{\city{York}, \postcode{YO10 5DD}, \country{United Kingdom}}}


\abstract{Measurement-device-independent quantum key distribution (MDI-QKD) was originally proposed as a means to address the issue of detector side-channel attacks and enable finite secure key rates over longer distances. However, the asymmetric characteristics of the channels from the two sources to the measurement device in MDI-QKD impose constraints on successfully extracting a secure key. In this work, we present a plug-and-play scheme for MDI-QKD based on differential phase shift (DPS) encoding. Specifically, we analyze the effects of pulse-width mismatch and polarization mismatch between the pulses arriving at the measurement device. The polarization mismatch is modeled with an assumption of sharing a common reference frame, and the maximum allowable mismatch is found to be $11$ degrees. Furthermore, we show that a channel length asymmetry of $176.5$ km results in Hong-Ou-Mandel interference visibility of $0.37$, thereby leading to zero secure key rates for a polarization-based MDI-QKD protocol. We then present a plug-and-play architecture for DPS-MDI-QKD as a solution to some of these issues, thereby paving the way for practical implementations of MDI protocols.}




\maketitle

\section{Introduction}\label{introduction}

Quantum key distribution (QKD) schemes establish identical and secure classical keys between two distant parties using quantum states. Fundamental features of quantum mechanics such as the uncertainty principle and the no-cloning theorem, offer the possibility of sharing unconditionally secure keys. These keys can then be used for symmetric key cryptography of classical information \cite{bennett2014}. The first QKD protocol presented in $1984$ \citep{bennett2014quantum}, the well known BB84 scheme, is based on encoding information into polarization states of light. This protocol essentially marked the birth of quantum cryptography and led to the development of other QKD schemes, for example, the B92 protocol \cite{bennett1992B92}, differential phase shift (DPS) QKD \cite{Inoue02, Inoue03}, coherent-one-way (COW) QKD \cite{stucki2005fast}, among others. Apart from this class of prepare-and-measure protocols, that use product states to establish the keys, QKD protocols relying on shared entanglement between two parties have also been proposed \cite{ekert1991quantum, bennett1992quantumBBM92, long2002theoretically}.\par 
In DPS-QKD, the transmitter sends a random bit string encoded as phase differences between consecutive weak coherent pulses in a pulse train. All the pulses generated from a laser source have the same phase reference within the coherence time of the laser. Differential phase encoding over such pulses can then be decoded using a delay line interferometer (DLI), and two detectors.\par
The unconditional security of point-to-point QKD protocols was originally proven based on fully characterized ideal single-photon sources and detectors \cite{10.1145/382780.382781, lo1999unconditional, shor2000simple, lo2005efficient}. Security proofs with practical weak coherent sources (WCSs) and detectors seek to establish an information-theoretic foundation for QKD \cite{10.5555/2011586.2011587, lo2005decoy}, but they still make assumptions about side-channel free sources and detectors. Most implementations use partially characterized devices, which opens them to side-channel attacks. Source imperfections such as incorrect state preparation, and the non-zero probability of multi-photon transmission, can be exploited by eavesdroppers to apply different attacks, such as the photon-number splitting (PNS) attack \cite{huttner1995quantum, brassard2000limitations}. Detector side-channel attacks such as detector blinding attack, fake state attack, and time shift attack \cite{makarov2009controlling,makarov2005faked,qi2005time}, expose the weakness in point-to-point QKD.\par
An alternate scheme that is free from side-channel attacks is device-independent (DI) QKD \cite{acin2007device}. In this scheme, key establishment relies on verifying the quantum behavior of the signal states using Bell inequalities \cite{bell1964einstein} and allows for the establishment of identical and secure keys without any device constraints. However, DI-QKD is unsuitable for a low-cost system implementation because it requires nearly unity detection efficiency and provides a very low key generation rate.
Measurement-device-independent (MDI) QKD was presented as an alternate, efficient solution to the problem of detector side-channel attacks \cite{lo2012measurement}. In this scheme, the two parties (Alice and Bob) who wish to establish a secure key, prepare states and send them to a third party (Charlie). The task of measurement is outsourced to Charlie, who could be an untrusted party. Charlie performs a joint measurement on the states received from Alice and Bob, and announces the measurement result. Based on the announced results, Alice and Bob sift their raw keys. Although Charlie could potentially know the relation (correlation or anti-correlation) between the sifted key bits of Alice and Bob, his uncertainty about the exact key bits is maximum. The original MDI-QKD protocol was based on polarization encoding. Subsequently, phase encoded MDI-QKD schemes are presented \cite{ma2012alternative, tamaki2012phase}. MDI and DI architecture have also been used for quantum secure direct communication (QSDC) application \cite{niu2018measurement, zhou2020device}. Silicon chip-based implementation of MDI-QKD is also explored in literature \cite{kwek2021chip}. More recently, an MDI-QKD scheme based on differential phase shift (DPS) encoding was presented \cite{ranu2021differential}. This scheme enjoys the benefits of both security against detector side-channel attacks, and robustness against phase distortion. 

%Our present work proposes an improved DPS-MDI-QKD sifting scheme, with a plug-and-play version better suited for field trials.\par
The secure key rate of MDI-QKD is a function of the visibility of Hong-Ou-Mandel (HOM) interference \cite{lo2012measurement, comandar2016quantum, da2013proof, tang2016measurement}. As single-photon sources are expensive and difficult to operate, WCSs are typically used in field implementations. The WCSs must have a good level of indistinguishability to provide high visibility in a HOM interference experiment. It is therefore critical to quantify the effects of various practical imperfections on the visibility of HOM interference. Analyzing the performance of the HOM interference between two independent WCSs ultimately helps in estimating realistic secure key rates of MDI-QKD implementations.\par
In past works, the effect of different factors such as the frequency difference between two independent pulses, detection-time deviation, spectral bandwidth, time jitter, the ratio of a beamsplitter (BS), unequal intensities in the two incident pulses, and afterpulsing of single-photon detectors (SPDs) on the visibility of HOM interference between two independent WCSs, have been quantified \cite{kim2014two, wang2017realistic, comandar2016near, da2015spectral, yuan2014interference}. Furthermore, it has been shown that the secure key rate of polarization-based MDI-QKD deteriorates due to poor HOM interference visibility, with the secure key rate dropping down to $0$ as soon as HOM interference visibility becomes less than $0.37$ \cite{xu2013practical, yuan2014interference}. However, the effect of pulse-shape mismatch, particularly pulse-width mismatch, on the visibility of HOM interference is missing in the literature. 

The effect of pulse-width mismatch becomes important in the context of time-bin-based MDI implementations \cite{ma2012alternative,ranu2021differential} and can be caused by many factors such as different behaviors of components in Alice and Bob's lab and channel length asymmetry between the two parties. In this work, we analyze the performance of MDI-QKD with a mismatch in pulse widths, and quantify the effect of channel length asymmetry on the secure key rates of the scheme. We also quantify the effect of polarization mismatch on the secure key rates of DPS-MDI-QKD.\par 
The original plug-and-play QKD protocol was proposed in 1997 \cite{muller1997plug} for point-to-point QKD schemes and then proven to be secure \cite{zhao2008quantum}. Subsequently, the plug-and-play scheme for MDI-QKD with an untrusted source was also proven to be secure \cite{xu2015measurement}. The Faraday mirror used in plug-and-play architecture mitigates birefringence effects and losses due to polarization rotation \cite{muller1997plug}. Thus in an ideal case, it provides the complete solution to the mismatch problem. Signals traveling to and fro through the same channel undergo a self-correction, making this architecture very attractive for implementation. Drawing motivation from these results, we propose a plug-and-play scheme for DPS-MDI-QKD and describe how this improves system performance in the context of pulse width and polarization mismatch.\par

The rest of the paper is organized as follows. We first review the DPS-MDI-QKD scheme in Sec. \ref{protocol}.  In Sec. \ref{modified_sifting}, we discuss an improved sifting scheme for DPS-MDI-QKD, which can readily be used for the plug-and-play version. The secure key rate of the DPS-MDI-QKD scheme with polarization mismatch is analyzed in Sec. \ref{pol_mismatch}. In Sec. \ref{pulse_mismatch}, we discuss the effect of channel asymmetry on pulse-width mismatch and the subsequent effect on the key rate of a polarization-based MDI protocol. Sec. $\ref{Pnp_protocol}$ comprises the plug-and-play version of the DPS-MDI-QKD protocol (considering a reliable source) and mitigation of the pulse width and polarization mismatch. Finally, we conclude with a summary in Sec. \ref{conclusion}. 

\section{DPS-MDI-QKD protocol } \label{protocol}
\begin{figure}[h]
    \centering
    \includegraphics[width=0.6\linewidth]{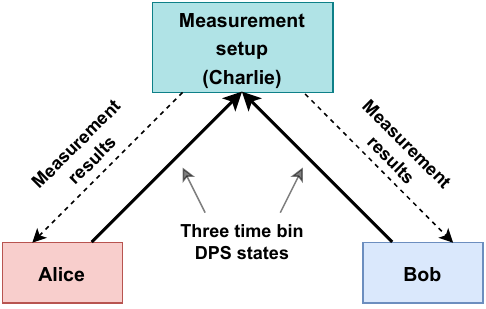}
    \caption{Block diagram for DPS-MDI-QKD.}
    \label{DPS_MDI}
\end{figure}

\begin{figure}[h]
    \centering
    \includegraphics[width=0.8\linewidth]{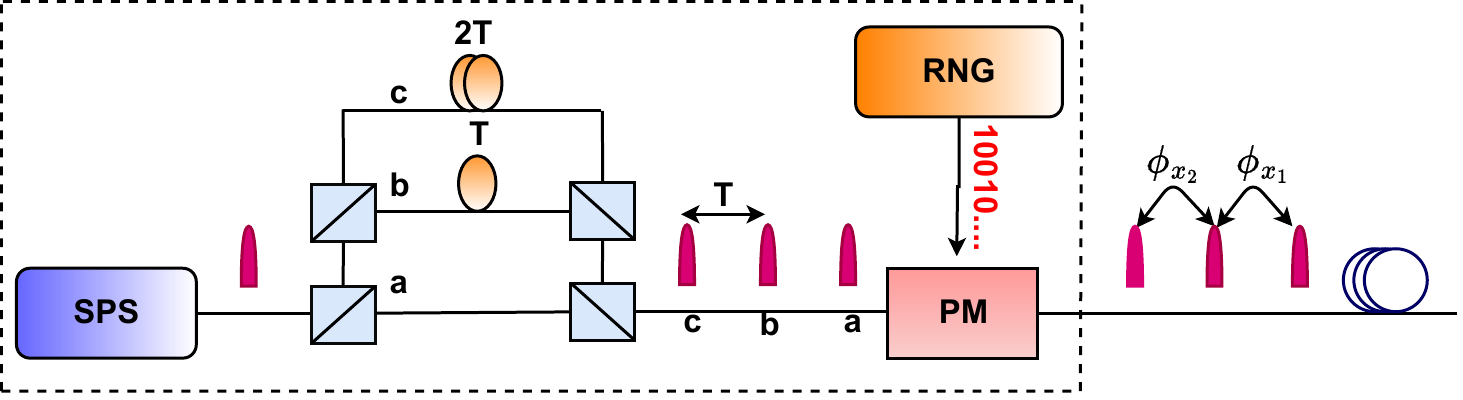}
    \caption{Schematic for preparing a differential phase encoded single-photon state in three-time bins. SPS: single-photon source, PM: phase modulator, RNG: random number generator}
    \label{state_prep}
\end{figure}

In the DPS-MDI-QKD protocol \cite{ranu2021differential}, Alice and Bob are two parties willing to establish identical and secure keys, as shown in Fig \ref{DPS_MDI}.
Both parties independently prepare single-photon states in a superposition of three-time bins using a three-path DLI, as shown in Fig. \ref{state_prep}. The superposition state at the output of the DLI is
\begin{equation}\label{sp_state}
    \ket{\psi_A}=\dfrac{1}{\sqrt{3}} \Big[|1\rangle_0 | 0\rangle_1 |0\rangle_2 + |0\rangle_0 |1\rangle_1 |0\rangle_2 + |0\rangle_0 |0\rangle_1|1\rangle_2\Big].
\end{equation}
Here $\ket{i}$ represents a Fock state with $i$ photons and the subscript represents the corresponding time bins. The time delay between consecutive time bins is the same. Differential phase encoding of information is done by introducing a phase difference between time bins. Two bits $x_1$ and $x_2$, generated by a random number generator (RNG), introduce the phase difference $\phi_{x_i} \in \{0,~\pi\}$ via a phase modulator (PM) as shown in Fig. 2. The corresponding output state is given by 
\begin{align}
\label{enc_state}
    %\ket{\psi}_{\text{enc}}=\frac{1}{\sqrt{3}}[\ket{100}_{t_1}+(-1)^{a_1}\ket{010}_{t_2}+(-1)^{a_2}\ket{001}_{t_3}]
    \ket{\psi_A}=\dfrac{1}{\sqrt{3}} \Big[|1\rangle_0 | 0\rangle_1 |0\rangle_2 +(-1)^{x_1} |0\rangle_0 |1\rangle_1 |0\rangle_2 
     +(-1)^{x_2} |0\rangle_0 |0\rangle_1 |1\rangle_2\Big].
\end{align}

Considering the two random bits generated by Alice (Bob) are $a_i$ ($b_i$) for $i\in \{1,2\}$. Bits $a_1$ ($b_1$) and $a_2$ ($b_2$) are encoded as the phase difference between the first-second and second-third time bins. We define 
\begin{align}
    \Delta\phi_1=\lvert\phi_{a1}-\phi_{b1}\rvert,\nonumber\\\Delta\phi_2=\lvert\phi_{a2}-\phi_{b2}\rvert,\nonumber
\end{align}
 where $\phi_{ai}$ and $\phi_{bi}$ are the phase value encoded based on the random bits $a_i$ and $b_i$. It is easy to verify that the quantity $\Delta\phi_1,~\Delta\phi_2~\in\{0,~\pi\}$.  Alice and Bob send these prepared states to a third party, Charlie, sitting at an equal distance from them. Charlie, who can be an eavesdropper, is given the responsibility of measuring the joint state received from Alice and Bob, and publicly announcing the results. Alice and Bob sift their keys based on these announcements. \par 

\subsection{Sifting scheme}
Charlie performs a Bell measurement on the joint state received from Alice and Bob. The measurement system in DPS-MDI-QKD ideally has two detectors (assume $c$ and $d$). Charlie announces the measurement results shown in Table \ref{sifting}. The listed outcome $(u,v,w)$ represents detector $u$ clicking in time bin 0, $v$ in time bin 1, and detector $w$ clicking in time bin 2 where $u,~v,~w~\in\{c,~d, ~0\}$. 0 in the outcome represents no detection in that time bin. Based on these results, Alice and Bob sift their keys. In our previous work \cite{ranu2021differential}, outcomes with single detections were not used for sifting and were discarded. Only the outcomes shown in Table \ref{sifting} with two clicks in two different time bins were used for sifting. The first four outcomes in Table \ref{sifting} assure the correlation or anti-correlation between the first bits used by Alice and Bob for encoding. The next four outcomes confirm the same with the second bits. The last four were discarded as no relation was found between the encoded phase. Thus a sifting rate of $\frac{4}{9}$ was achieved.
\begin{table}[h]
\caption{Sifting scheme for DPS-MDI-QKD \cite{ranu2021differential}. Here outcome $(u,v,w)$ represents detector $u$ clicking in time bin 0 and $v$ in time bin 1, and detector $w$ clicking in time bin 2 where $u,~v,~w~\in\{c,~d, ~0\}$. 0 in the outcome represents no detection in that time bin. Also $\overline{a}$ represents the complement of bit $a$.}\label{sifting}%
\begin{tabular}{@{}lcccc@{}}
\toprule
\textbf{S. No.} & \textbf{Outcomes} & \textbf{Phase} & \textbf{Alice's} & \textbf{Bob's} \\ 
        &           & \textbf{relation}    & \textbf{key}    & \textbf{key}\\
\midrule
1 & $(c,c,0)$ & \multirow{2}{*}{$\Delta\phi_1=0$} & \multirow{2}{*}{$a_1$} & \multirow{2}{*}{$b_1$} \\ 
 2 & $(d,d,0)$ &  &  & \\
 3 & $(c,d,0)$ & \multirow{2}{*}{$\Delta\phi_1=\pi$} & \multirow{2}{*}{$a_1$} & \multirow{2}{*}{$\overline{b_1}$} \\ 
 4 & $(d,c,0)$ &  &  & \\
 \hline
 5 & $(c,0,c)$ & \multirow{2}{*}{$\Delta\phi_2=0$} & \multirow{2}{*}{$a_2$} & \multirow{2}{*}{$b_2$} \\ 
 6 & $(d,0,d)$ &  &  & \\
 7 & $(c,0,d)$ & \multirow{2}{*}{$\Delta\phi_2=\pi$} & \multirow{2}{*}{$a_2$} & \multirow{2}{*}{$\overline{b_2}$} \\ 
 8 & $(d,0,c)$ &  &  & \\
 \hline
 9 & $(0,c,c)$ & \multirow{4}{*}{-} & \multirow{4}{*}{-} & \multirow{4}{*}{-} \\ 
 10 & $(0,d,d)$ &  &  & \\
 11 & $(0,c,d)$ & & &  \\ 
 12 & $(0,d,c)$ &  &  & \\
\botrule
\end{tabular}
\end{table}
\section{Improved sifting scheme for DPS-MDI-QKD}\label{modified_sifting}
In this section, we discuss a revised sifting scheme, that increases the sifting factor of DPS-MDI-QKD and improves the secure key rate. Here, we show that the four outcomes excluded in Table \ref{sifting} can also be used for sifting. Including these outcomes improves the sifting rate to $\frac{2}{3}$.\par
The four remaining outcomes that we use for sifting along with the previous scheme are $(0,c,c),~(0,d,d), ~(0,c,d)$, and $(0,d,c)$. If we see the state after interference at Charlie \cite{ranu2021differential}, we quickly realize that the output states $(0,c,c)$ and $(0,d,d)$ imply that either $\Delta\phi_{1}=\Delta\phi_{2}=0$ or $\Delta\phi_{1}=\Delta\phi_{2}=\pi$ which assure that $\Delta\phi_{1}=\Delta\phi_{2}$. Similarly output states $(0,c,d)$ and $(0,d,c)$  assure that $\lvert\Delta\phi_{1}-\Delta\phi_{2}\rvert=\pi$. Here, we use these relations to find the correlation or anti-correlation between the encoded bits of Alice and Bob and use them to improve the sifting rate.\par
 As discussed earlier, logic 0 and logic 1 are encoded as a phase difference of 0 and $\pi$, respectively. Therefore $\Delta\phi_1$ is equal to 0 or $\pi$ implies that $a_1\oplus b_1$ is 0 or 1, respectively. Similarly, $\Delta\phi_2$ is equal to 0 or $\pi$ implies that $a_2\oplus b_2$ is 0 or 1. Therefore the announcement of the outcomes $(0,c,c)$ or $(0,d,d)$ assure Alice and Bob that
\begin{align}
    \Delta\phi_{1}&=\Delta\phi_{2}\nonumber\\
    \implies a_1\oplus b_1&= a_2\oplus b_2\nonumber\\
    \implies a_1\oplus a_2&=b_1\oplus b_2.
\end{align}
Therefore, no individual bit of Alice and Bob can be directly correlated for these outcomes, but the XOR of both the bits of Alice is correlated to the XOR of both the bits of Bob. Similarly, there is an anti-correlation between the XOR of the two bits encoded in the remaining outcomes.
\begin{align}
    &\lvert\Delta\phi_{1}-\Delta\phi_{2}\rvert=\pi\nonumber\\
    \implies &(a_1\oplus a_2) \oplus (b_1\oplus b_2)=1\nonumber\\
    \implies &\overline{(a_1\oplus a_2)}=b_1\oplus b_2.
\end{align}
The complete sifting scheme is shown in Table \ref{mod_sifting}.\par
\begin{table}[h]
\caption{Improved sifting scheme for DPS-MDI-QKD.}\label{mod_sifting}%
\begin{tabular}{@{}lcccc@{}}
\toprule
\textbf{S. No.} & \textbf{Outcomes} & \textbf{Phase} & \textbf{Alice's} & \textbf{Bob's} \\ 
        &           & \textbf{relation}    & \textbf{key}    & \textbf{key}\\
\midrule
1 & $(c,c,0)$ & \multirow{2}{*}{$\Delta\phi_1=0$} & \multirow{2}{*}{$a_1$} & \multirow{2}{*}{$b_1$} \\ 
 2 & $(d,d,0)$ &  &  & \\
 3 & $(c,d,0)$ & \multirow{2}{*}{$\Delta\phi_1=\pi$} & \multirow{2}{*}{$a_1$} & \multirow{2}{*}{$\overline{b_1}$} \\ 
 4 & $(d,c,0)$ &  &  & \\
 \hline
 5 & $(c,0,c)$ & \multirow{2}{*}{$\Delta\phi_2=0$} & \multirow{2}{*}{$a_2$} & \multirow{2}{*}{$b_2$} \\ 
 6 & $(d,0,d)$ &  &  & \\
 7 & $(c,0,d)$ & \multirow{2}{*}{$\Delta\phi_2=\pi$} & \multirow{2}{*}{$a_2$} & \multirow{2}{*}{$\overline{b_2}$} \\ 
 8 & $(d,0,c)$ &  &  & \\
 \hline
 9 & $(0,c,c)$ & \multirow{2}{*}{$\lvert\Delta\phi_1-\Delta\phi_2\rvert=0$} & \multirow{2}{*}{$a_1\oplus a_2$} & \multirow{2}{*}{$b_1\oplus b_2$} \\ 
 10 & $(0,d,d)$ &  &  & \\
 11 & $(0,c,d)$ & \multirow{2}{*}{$\lvert\Delta\phi_1-\Delta\phi_2\rvert=\pi$}& \multirow{2}{*}{$a_1\oplus a_2$} & \multirow{2}{*}{$\overline{b_1\oplus b_2}$}  \\ 
 12 & $(0,d,c)$ &  &  & \\
\botrule
\end{tabular}
\end{table}
Now, we discuss the secure key rate with ideal single-photon sources in DPS-MDI-QKD, with the improved sifting scheme discussed in Table \ref{mod_sifting}. The secure key rate, in this case, is given by
\begin{equation}\label{single_rate}
    R=Y_{11}\Big[1-f\cdot H(e_b)-H(e_p)\Big],
\end{equation}
where $Y_{11}$ is the yield due to single photons, defined as the probability of making a successful detection given both Alice and Bob are sending single photons. $e_b$ and $e_p$ are bit and phase error rates, and $e_p$ is upper bounded by $e_b$ \cite{ranu2021differential}. $f$ is error correction inefficiency and $$H(x)=-x\text{log}_2(x)-(1-x)\text{log}_2(1-x)$$ is the Shannon entropy of $x$. The yield for the outcome $(c,c,0)$, $Y_{11}^{(c,c,0)}$ is given by
\begin{multline}\label{yield}
    Y_{11}^{(c,c,0)}=(1-P_{\text{dark}})^4\Bigg[\frac{\eta_a\eta_b}{18}+P_{\text{dark}}\Bigg(\frac{\eta_a+\eta_b}{3}-\frac{5\eta_a\eta_b}{9}\Bigg)\\+P_{\text{dark}}^2(1-\eta_a)(1-\eta_b)\Bigg].
\end{multline}
Here, $P_{\text{dark}}$ is the dark count rate per pulse, and $\eta_a$ and $\eta_b$ are the channel transmittance from Alice and Bob, respectively, to Charlie. It is easy to verify that the yield for all 12 outcomes discussed in Table \ref{mod_sifting} is the same as given in Eq. (\ref{yield}). Therefore, the overall yield is
\begin{equation}
    Y_{11}=12Y_{11}^{(c,c,0)}.    
\end{equation}
The bit error rate $e_b$ is defined as the probability of announcement of an outcome leading to the correlation between bits of Alice and Bob, given that they sent anti-correlated bits, or vice versa. e.g. announcement of $(c,c,0)$ or $(d,d,0)$ when $\Delta\phi_1=\pi$. From the yield expression in Eq. (\ref{yield}), it is clear that the error is possible due to dark counts. Therefore, the bit error rate is 
\begin{multline}
    e_{b}^{\prime}Y_{11}=12(1-P_{\text{dark}})^4\Bigg[P_{\text{dark}}\Bigg(\frac{\eta_a+\eta_b}{3}-\frac{5\eta_a\eta_b}{9}\Bigg)+P_{\text{dark}}^2(1-\eta_a)(1-\eta_b)\Bigg].
\end{multline}
If we consider the phase misalignment error $e_d$, there will be an error even when both the photons reach Charlie. The bit error rate in this case, is
\begin{multline}\label{s_eb}
    e_{b}Y_{11}=12(1-P_{\text{dark}})^4\Bigg[e_d\frac{\eta_a\eta_b}{18}+P_{\text{dark}}\Bigg(\frac{\eta_a+\eta_b}{3}-\frac{5\eta_a\eta_b}{9}\Bigg)+\\P_{\text{dark}}^2(1-\eta_a)(1-\eta_b)\Bigg].
\end{multline}
\begin{figure}
        \centering
        \includegraphics[width=0.6\linewidth]{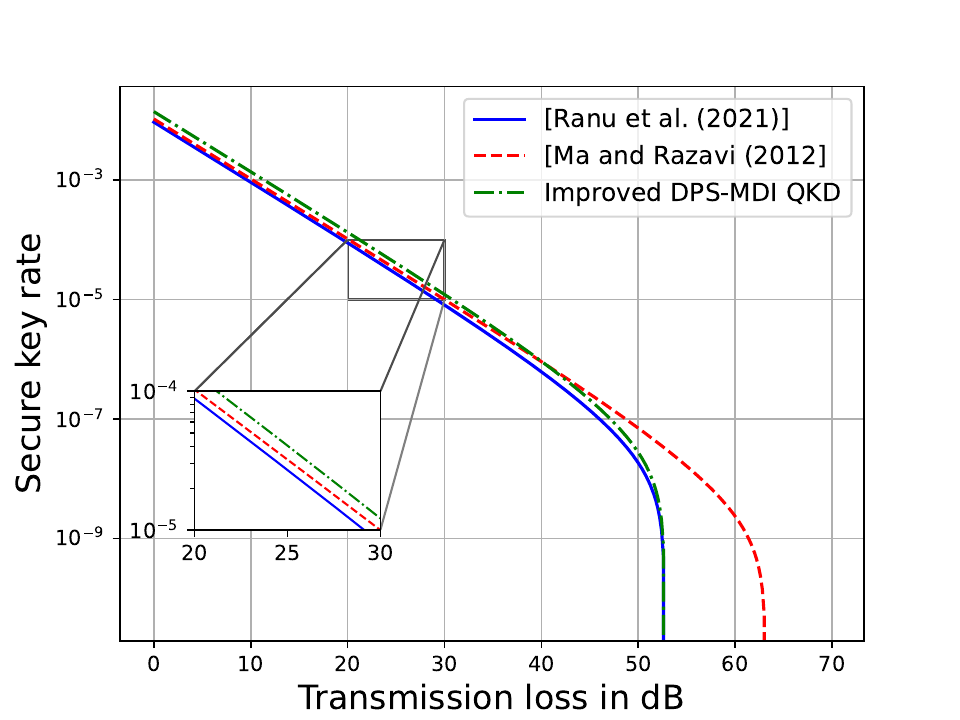}
        \caption{Comparison of secure key rate of DPS-MDI-QKD without and with improved sifting scheme. The key rates are derived considering single-photon sources in all the cases. Parameters used for simulations such as dark count probability, attenuation, and detector efficiency are considered the same as those used for the simulation of DPS-MDI-QKD key rate\cite{ranu2021differential}.}
        \label{mod_scheme}
    \end{figure}
    %Similar to the yield, the bit error rate for all the useful outcomes is the same as shown in the expression in Eq. (\ref{s_eb}). Therefore, the overall bit error rate for the scheme is
%\begin{equation}\label{error_rate}
%    e_bY_{11}=12e_{b}^{(c,c,0)}Y_{11}.
%\end{equation}
    Using the derived bit error rate and single-photon yield in Eq. (\ref{single_rate}), we derive and plot the secure key rate of the improved scheme against the original sifting scheme. The result is shown in Fig. \ref{mod_scheme}, where we have also compared the secure key rates of \cite{ranu2021differential} with and without improved sifting against phase-based MDI-QKD \cite{ma2012alternative}. We found that the original scheme was inferior to the phase-based MDI-QKD in both the key rate and the maximum distance up to which keys can be established. But with improved sifting, the key rate of DPS-MDI is not only better than \cite{ranu2021differential} but is also better than the phase based MDI-QKD \cite{ma2012alternative} up to 40 dB channel loss. It is clear that the improvement in the sifting ratio from $\frac{4}{9}$ to $\frac{2}{3}$ leads to an improvement in the secure key rate of DPS-MDI-QKD.
\section{Secure key rates with polarization mismatch}\label{pol_mismatch}
As discussed in previous works, the channel asymmetry in the implementation of MDI-QKD schemes leads to a reduction in the visibility of HOM interference~\cite{kim2014two, wang2017realistic, comandar2016near, da2015spectral, yuan2014interference}, which is a bottleneck for the performance of such QKD schemes. There can be many factors responsible for this asymmetry, such as physical lengths of the fibers, external environment, birefringence, dispersion, components used, etc.  In this section, we study in detail the effect of polarization mismatch due to channel asymmetry on the performance of DPS-MDI-QKD. Polarization mismatch arises due to the single-mode fibers used in terrestrial implementations, due to the birefringence and the effect of the external environment. This affects the secure key rates of DPS-MDI-QKD. Here, we consider the decoy state implementation of the protocol, and assume that the states prepared by the two parties are linearly polarized. \par 
\subsection{Overall gain and quantum bit error rate (QBER)}
In DPS-MDI-QKD, Alice (Bob) splits a pulse with mean photon number $\mu_a$ ($\mu_b$) into three pulses of intensity $\mu_a/3$ ($\mu_b/3$) each. The subscript $a$ and $b$ represent Alice and Bob throughout the paper. All the weak coherent states used here are supposed to be phase-randomized to make them immune to the photon-number splitting attack. This makes the states equivalent to a classical mixture of Fock states. Assuming the symmetric configuration in which Alice and Bob are at an equal distance from Charlie, the transmittance of the channel from Alice to Charlie ($\eta_a$) and Bob to Charlie ($\eta_b$) is given by 
\begin{align}\label{eq9}
    \eta_a=\eta_b=\eta_{\text{det}}10^{-\gamma l/20},
\end{align}
where $l$ is the distance between Alice and Bob. The detection efficiencies of the two detectors at the two output ports of the beamsplitter are considered equal and given by $\eta_{\text{det}}$, and $\gamma$ is the attenuation of the channel in dB/km. Therefore, the effective mean photon number of the pulses reaching the beamsplitter, in each time bin, is $\frac{\eta_a\mu_a}{3}$ and $\frac{\eta_b\mu_b}{3}$. We consider that the pulses are identical in all the degrees of freedom except polarization. Hence, if the unit field vector showing the polarization of Alice and Bob are $\hat{e_a}$ and $\hat{e_b}$ respectively, and if they have a polarization mismatch of angle $\Phi$ between them, then $\hat{e_a}\cdot\hat{e_b}=\text{cos}(\Phi)$. We represent the two unit vectors in terms of the components along horizontal $\hat{e_h}$ and vertical $\hat{e_v}$ polarization direction as \cite{moschandreou2018experimental}
\begin{align}\label{eq10}
\hat{e_a}&=\hat{e_a}\cdot\hat{e_h}~\hat{e_h}+\hat{e_a}\cdot\hat{e_v}~\hat{e_v}\nonumber,\\
    \hat{e_b}&=\hat{e_b}\cdot\hat{e_h}~\hat{e_h}+\hat{e_b}\cdot\hat{e_v}~\hat{e_v}.
\end{align}
We also assume that the two lasers, one with Alice and the other with Bob are phase-locked. Therefore, the input state of the beamsplitter is
\begin{equation}\label{eq11}
    \ket{\psi_{\text{in}}}= \ket{\alpha}_{0}\ket{\alpha e^{i\phi_{a_1}}}_{1}\ket{\alpha e^{i\phi_{a_2}}}_{2}\ket{\beta}_{0}\ket{\beta e^{i\phi_{b_1}}}_{1}\ket{\beta e^{i\phi_{b_2}}}_{2},
\end{equation}
where $\alpha=\lvert\alpha\rvert e^{i\theta_a}$ and $\beta=\lvert\beta\rvert e^{i\theta_b}$. The quantities $\theta_a$ and $\theta_b$ are the randomized phase values chosen by Alice and Bob. Similarly $\phi_{x_i}$ is a differential phase encoded by $x\in\{a,b\}$ and $i\in \{1,2\}$. Here, $a$ and $b$ represent Alice and Bob, respectively. Without loss of generality, we consider the input state to the beamsplitter at time bin 0 as,
\begin{equation}\label{eq12}
    \ket{\psi_{\text{in}}}_0=\ket{\alpha}_{0}\ket{\beta }_{0}.
\end{equation}
We know that a coherent state can be expressed as, 
\begin{equation}\label{eq13}
    \ket{\alpha}= e^\frac{-\lvert{\alpha}\rvert^2}{2}e^{\alpha a^{\dagger}}\ket{0},
\end{equation}
where $a^{\dagger}$ is the creation operator for the state. We can write the input state at time bin 0 in terms of the operator as
\begin{equation}\label{eq14}
    \ket{\psi_{\text{in}}}_0=e^\frac{-(\lvert{\alpha}\rvert^2+\lvert{\beta}\rvert^2)}{2}e^{\alpha a^{\dagger}}e^{\beta b^{\dagger}}\ket{0},
\end{equation}
\begin{figure}[h]
    \centering
    \includegraphics[width=0.5\linewidth]{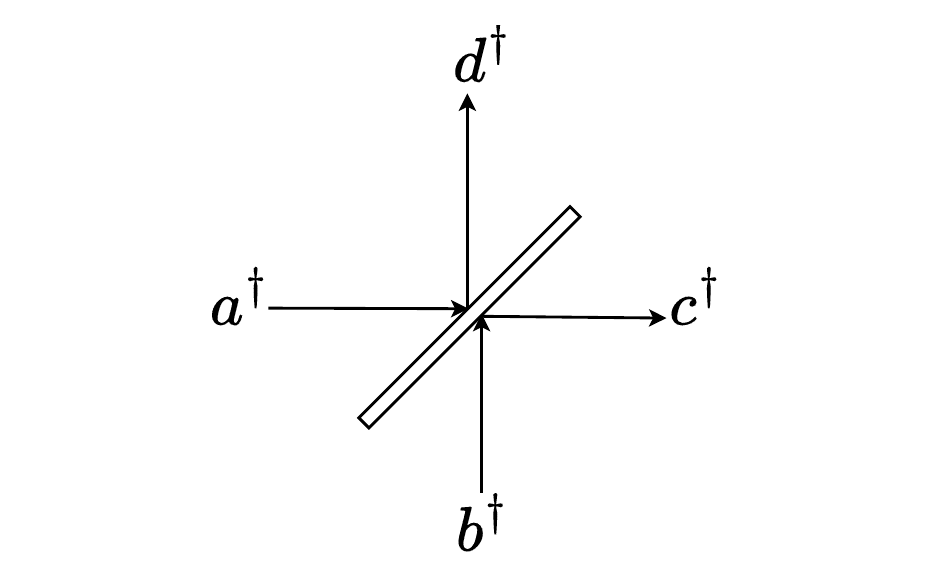}
    \caption{50:50 beamsplitter used in the measurement setup of Charlie.}
    \label{BS}
\end{figure}
where $a^{\dagger},~b^{\dagger} $ are the creation operators at input ports $a$ and $b$ of the beamsplitter shown in Fig. \ref{BS}. These operators can be represented as a superposition of creation operators along horizontal and vertical polarization as \cite{moschandreou2018experimental}
\begin{align}\label{eq15}
a^{\dagger}&=\hat{e_a}\cdot\hat{e_h}~a_h^{\dagger}+\hat{e_a}\cdot\hat{e_v}~a_v^{\dagger},\nonumber\\
b^{\dagger}&=\hat{e_b}\cdot\hat{e_h}~b_h^{\dagger}+\hat{e_b}\cdot\hat{e_v}~b_v^{\dagger},
\end{align}
where $a_h^{\dagger}$ is the creation operator for a mode with horizontal polarization at port $a$. We similarly define the other necessary operators. The beamsplitter transforms the input state as
 \begin{align}\label{eq17}
     a^{\dagger}&\rightarrow\frac{1}{\sqrt{2}}[c^{\dagger}+d^{\dagger}],\nonumber\\
     b^{\dagger}&\rightarrow\frac{1}{\sqrt{2}}[c^{\dagger}-d^{\dagger}].
 \end{align}
% \end{multicols}
 Applying the transformation to the input state at time bin 0, we get
 \begin{align}\label{eq18}
\ket{\psi_{\text{out}}}_{0}&=\ket{\alpha_{ch}}\ket{\alpha_{cv}}\ket{\alpha_{dh}}\ket{\alpha_{dv}},
 \end{align}
  where
  \begin{align}
      {\alpha_{ch}}&={\alpha\hat{e_a}\cdot\hat{e_h}+\beta\hat{e_b}\cdot\hat{e_h}},\hspace{0.7cm}{\alpha_{cv}}={\alpha\hat{e_a}\cdot\hat{e_v}+\beta\hat{e_b}\cdot\hat{e_v}},\nonumber\\
      {\alpha_{dh}}&={\alpha\hat{e_a}\cdot\hat{e_h}-\beta\hat{e_b}\cdot\hat{e_h}},\hspace{0.7cm}
      {\alpha_{dv}}={\alpha\hat{e_a}\cdot\hat{e_v}-\beta\hat{e_b}\cdot\hat{e_v}}.\nonumber
  \end{align} 
  Here $\lvert\alpha_{ch}\rvert^2$ represents the mean photon number of the coherent states $\ket{\alpha_{ch}}$ at port $c$ with polarization $h$. Other states are defined similarly. Using the output modes, we derive the probability of getting $m$ photons at port $c$ as 
\begin{align}\label{eq21}
    P_m^c&=\sum_{m_h=0}^m P_{m_h}^{c_h}\times P_{m-m_h}^{c_v}\nonumber\\
    &=\frac{e^{-\lvert\alpha_{ch}\rvert^2-\lvert\alpha_{cv}\rvert^2}}{m!}\left(\lvert\alpha_{ch}\rvert^2+\lvert\alpha_{cv}\rvert^2\right)^m.
\end{align}
We can also define $\mu_{c0}=\lvert\alpha_{ch}\rvert^2+\lvert\alpha_{cv}\rvert^2$ as the resultant mean photon number at port $c$ at time bin 0. Thus the probability of getting a detection at port $c$ in this time bin is
\begin{equation}\label{eq22}
    P_{c0}=1-\left(1-P_{\text{dark}}\right)\times e^{-\mu_{c0}}.
\end{equation}
Similarly, we derive the probability of getting $m$ photons at port $d$ as
\begin{align}\label{eq23}
     P_m^d&=\frac{e^{-\lvert\alpha_{dh}\rvert^2-\lvert\alpha_{dv}\rvert^2}}{m!}\left(\lvert\alpha_{dh}\rvert^2+\lvert\alpha_{dv}\rvert^2\right)^m,
\end{align}
and define $\mu_{d0}=\lvert\alpha_{dh}\rvert^2+\lvert\alpha_{dv}\rvert^2$ as the resultant mean photon number at port $d$. Thus the probability of detection at port $d$ is
\begin{equation}\label{eq24}
    P_{d0}=1-(1-P_{\text{dark}})\times e^{-\mu_{d0}}.
\end{equation}
Following the same approach, we can find the probabilities for detectors $c$ and $d$ clicking in the other two time bins. These probabilities are
\begin{align}\label{eq28}
    P_{c1}&=1-(1-P_{\text{dark}})\times e^{-\mu_{c1}}, \hspace{1cm}
    P_{c2}=1-(1-P_{\text{dark}})\times e^{-\mu_{c2}},\nonumber\\
    P_{d1}&=1-(1-P_{\text{dark}})\times e^{-\mu_{d1}}, \hspace{1cm}
    P_{d2}=1-(1-P_{\text{dark}})\times e^{-\mu_{d2}},
\end{align}
where the mean photon numbers at port $c~\text{and }d$, at time bin $i \in \{1,2\}$, are
\begin{equation}\label{eq29}
    \mu_{ci}=\frac{1}{2}\left[\lvert\alpha\rvert^2+\lvert\beta\rvert^2+2\lvert\alpha\rvert\lvert\beta\rvert\text{cos}(\delta)\text{cos}(\Phi)\right],\nonumber
\end{equation}
\begin{equation}
    \mu_{di}=\frac{1}{2}\left[\lvert\alpha\rvert^2+\lvert\beta\rvert^2-2\lvert\alpha\rvert\lvert\beta\rvert\text{cos}(\delta)\text{cos}(\Phi)\right],
\end{equation}
and $\delta=\theta_a-\theta_b+\phi_{ai}-\phi_{bi}$. As shown in Table \ref{sifting}, there are eight useful outcomes for which a correlation or anti-correlation exists in Alice and Bob's bits. We use the probabilities of various detectors clicking at different time bins to find the probability of successful outcomes and derive the gain. The overall gain $Q_{\mu_a, \mu_b}$ is defined as the probability of successful detection and is given by 
\begin{align}\label{eq30}
Q_{\mu_a,\mu_b}=P_{\Delta\phi_1=0}^{(c,c,0)}+P_{\Delta\phi_1=0}^{(d,d,0)}+P_{\Delta\phi_1=\pi}^{(c,d,0)}+P_{\Delta\phi_1=\pi}^{(d,c,0)}+P_{\Delta\phi_2=0}^{(c,0,c)}+\nonumber\\P_{\Delta\phi_2=0}^{(d,0,d)}+P_{\Delta\phi_2=\pi}^{(c,0,d)}+P_{\Delta\phi_2=\pi}^{(d,0,c)}\nonumber\\
    =4y^4\Big[\zeta^2+\zeta^{-2}-2y\zeta-2y\zeta^{-1}+2y^2\Big],
\end{align}
where $P_{\Delta\phi_1=0}^{(c,c,0)}$ is the probability of getting only detector $c$ clicking at time bin 0 and 1, when $\Delta\phi_1=0$. Similarly the other terms are defined. Here $\zeta=e^{x\text{cos}(\Delta\theta)\text{cos}(\Phi)}$, $y=(1-p_{\text{dark}})e^{-\mu'/6}$, $x=\frac{\sqrt{\eta_a\mu_a\eta_b\mu_b}}{3}$, and $\mu'=\eta_a\mu_a+\eta_b\mu_b$. Considering key sifting only when Alice and Bob use exactly the same random phase gives zero key rates. Therefore, a more practical treatment is to divide the phase plane into $N$ phase slices of width $\frac{2\pi}{N}$. Key sifting is done whenever Alice and Bob choose phase values within the same phase slice. We note that $N=16$ gives the optimal key rate \cite{ranu2021differential}. The corresponding gain and QBER are given by
\begin{multline}\label{overgain}
   Q_m=\frac{N}{\pi^2}\int_{\theta_b=0}^{\pi/N}\int_{\theta_a=m\pi/N}^{(m+1)\pi/N}4y^4\Big[\zeta^2+\zeta^{-2}-2y\zeta-2y\zeta^{-1}+2y^2\Big]d\theta_bd\theta_a,
\end{multline}
and
\begin{multline}\label{overqber}
    E_mQ_m=\frac{N}{\pi^2}\int_{\theta_b=0}^{\pi/N}\int_{\theta_a=m\pi/N}^{(m+1)\pi/N}8y^4\left[1-y\zeta-y\zeta^{-1}+y^2\right]d\theta_bd\theta_a.
\end{multline} 
%Here, we consider that Bob always chooses a phase value from the first slice, and Alice can choose any value.
 Eqs. (\ref{overgain}) and (\ref{overqber}) represent the effect of polarization mismatch on the overall gain and QBER for DPS-MDI-QKD using WCSs. To estimate the secure key rate of the scheme, we must also find the gain and QBER with single-photon sources. 
\subsection{Single-photon gain and QBER}
\begin{figure}[htbp]
    \centering
    \includegraphics[width=0.6\linewidth]{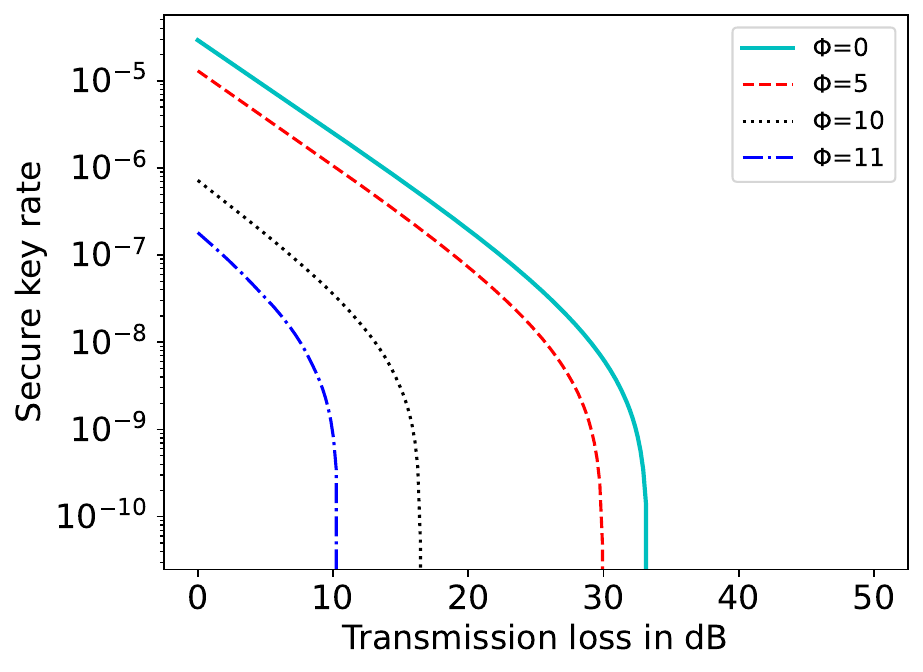}
    \caption{Secure key rate for DPS-MDI-QKD with polarization mismatch $\Phi$.}
    \label{pol_mism}
\end{figure}

We estimate the gain and QBER in the polarization mismatch scenario when Alice and Bob use single-photon sources. The joint state of Alice and Bob sending single photons in DPS-MDI-QKD is
\begin{multline} \label{eq35}
    \ket{\psi_{\text{in}}}=\frac{1}{3}\Big[a_0^{\dagger}+e^{i\phi_{a1}}a_1^{\dagger}+e^{i\phi_{a2}}a_2^{\dagger}\Big]\otimes\Big[b_0^{\dagger}+e^{i\phi_{b1}}b_1^{\dagger}+e^{i\phi_{b2}}b_2^{\dagger}\Big]\ket{00}_{ab}.
\end{multline}
Here $a$ or $b$ represent the input modes to the beamsplitter from Alice and Bob respectively, and $i\in~\{0,1,2\}$ represents the time bin. Assuming that the polarization of mode $a$ ($b$) is linear, and makes an angle $\theta_a$ ($\theta_b$) with respect to the horizontal direction, we represent the creation operators of the input modes as
\begin{align}\label{eq36}
    a^{\dagger}&=\text{cos}(\theta_a)~a_h^{\dagger}+\text{sin}(\theta_a)~a_v^{\dagger},\nonumber\\
    b^{\dagger}&=\text{cos}(\theta_b)~b_h^{\dagger}+\text{sin}(\theta_b)~b_v^{\dagger}.
\end{align}
 Based on the transmittance $\eta_a$ ($\eta_b$) from Alice (Bob) to Charlie, the state reaching Charlie is a mixed state given by,
 \begin{multline}\label{eq38}
     \rho_{\text{in}}=\frac{\eta_a\eta_b}{9}\ket{\psi_{\text{11}}}\bra{\psi_{\text{11}}}+\frac{\eta_a(1-\eta_b)}{3}\ket{\psi_{\text{10}}}\bra{\psi_{\text{10}}}+\frac{\eta_b(1-\eta_a)}{3}\ket{\psi_{\text{01}}}\bra{\psi_{\text{01}}}\\+(1-\eta_a)(1-\eta_b)\ket{\psi_{{\text{00}}}}\bra{\psi_{{\text{00}}}}.
 \end{multline}
 Here $\ket{\psi_{\text{11}}}$ is the state when photons from both Alice and Bob reach Charlie. Similarly, $\ket{\psi_{\text{10}}}$ ($\ket{\psi_{\text{01}}}$) represents the state when a photon only from Alice (Bob) reaches Charlie. $\ket{\psi_{\text{00}}}$ is the state when no photon reaches Charlie. Using Eqs. (\ref{eq35}) and (\ref{eq36}) we find the state $\ket{\psi_{\text{11}}}$ at the beamsplitter with a mismatch in polarization. The beamsplitter transforms such an input state, and the corresponding output state is
\begin{align}\label{eq39}
    \ket{\psi_{\text{11}}^{\text{out}}}=&\frac{1}{2}\Big[\Big(\text{cos}(\theta_a)~(c_{\text{h0}}^{\dagger}+d_{\text{h0}}^{\dagger})+\text{sin}(\theta_a)~(c_{\text{v0}}^{\dagger}+d_{\text{v0}}^{\dagger})\Big)+\nonumber\\&e^{i\phi_{a1}}\Big(\text{cos}(\theta_a)~(c_{\text{h1}}^{\dagger}+d_{\text{h1}}^{\dagger})+\text{sin}(\theta_a)~(c_{\text{v1}}^{\dagger}+d_{\text{v1}}^{\dagger})\Big)+\nonumber\\&e^{i\phi_{a2}}\Big(\text{cos}(\theta_a)~(c_{\text{h2}}^{\dagger}+d_{\text{h2}}^{\dagger})+\text{sin}(\theta_a)~(c_{\text{v2}}^{\dagger}+d_{\text{v2}}^{\dagger})\Big)\Big]\nonumber\\&\otimes\Big[\Big(\text{cos}(\theta_b)~(c_{\text{h0}}^{\dagger}-d_{\text{h0}}^{\dagger})+\text{sin}(\theta_b)~(c_{\text{v0}}^{\dagger}-d_{\text{v0}}^{\dagger})\Big)+\nonumber\\&e^{i\phi_{b1}}\Big(\text{cos}(\theta_b)~(c_{\text{h1}}^{\dagger}-d_{\text{h1}}^{\dagger})+\text{sin}(\theta_b)~(c_{\text{v1}}^{\dagger}-d_{\text{v1}}^{\dagger})\Big)+\nonumber\\&e^{i\phi_{b2}}\Big(\text{cos}(\theta_b)~(c_{\text{h1}}^{\dagger}-d_{\text{h2}}^{\dagger})+\text{sin}(\theta_b)~(c_{\text{v2}}^{\dagger}-d_{\text{v2}}^{\dagger})\Big)\Big]\ket{00}_{cd}.
\end{align}
Similarly, we can write the output state of the beamsplitter corresponding to the other three possible input states as,
\begin{align}
    \ket{\psi_{\text{00}}^{\text{out}}}=&\ket{00}_{cd},\nonumber\\
    \ket{\psi_{\text{10}}^{\text{out}}}=&\frac{1}{\sqrt{2}}\Big[\Big(\text{cos}(\theta_a)~(c_{\text{h0}}^{\dagger}+d_{\text{h0}}^{\dagger})+\text{sin}(\theta_a)~(c_{\text{v0}}^{\dagger}+d_{\text{v0}}^{\dagger})\Big)+\nonumber\\&e^{i\phi_{a1}}\Big(\text{cos}(\theta_a)~(c_{\text{h1}}^{\dagger}+d_{\text{h1}}^{\dagger})+\text{sin}(\theta_a)~(c_{\text{v1}}^{\dagger}+d_{\text{v1}}^{\dagger})\Big)+\nonumber\\&e^{i\phi_{a2}}\Big(\text{cos}(\theta_a)~(c_{\text{h2}}^{\dagger}+d_{\text{h2}}^{\dagger})+\text{sin}(\theta_a)~(c_{\text{v2}}^{\dagger}+d_{\text{v2}}^{\dagger})\Big)\Big]\ket{00}_{cd},\nonumber
    \end{align}
    \begin{align}\label{eq41}
    \ket{\psi_{\text{01}}^{\text{out}}}=&\frac{1}{\sqrt{2}}\Big[\Big(\text{cos}(\theta_b)~(c_{\text{h0}}^{\dagger}-d_{\text{h0}}^{\dagger})+\text{sin}(\theta_b)~(c_{\text{v0}}^{\dagger}-d_{\text{v0}}^{\dagger})\Big)+\nonumber\\&e^{i\phi_{b1}}\Big(\text{cos}(\theta_b)~(c_{\text{h1}}^{\dagger}-d_{\text{h1}}^{\dagger})+\text{sin}(\theta_b)~(c_{\text{v1}}^{\dagger}-d_{\text{v1}}^{\dagger})\Big)+\nonumber\\&e^{i\phi_{b2}}\Big(\text{cos}(\theta_b)~(c_{\text{h1}}^{\dagger}-d_{\text{h2}}^{\dagger})+\text{sin}(\theta_b)~(c_{\text{v2}}^{\dagger}-d_{\text{v2}}^{\dagger})\Big)\Big]\ket{00}_{cd}.
\end{align}
Using the above states, we derive the yield $Y_{11}^{(c,c,0))}$ as 
\begin{multline}\label{eq42}
    Y_{11}^{(c,c,0)}=(1-P_{\text{\text{dark}}})^4\Bigg[\frac{\eta_a\eta_b}{18}\times \left(1+\text{cos}^2(\Phi)\right)+P_{\text{dark}}\Bigg(\frac{\eta_a+\eta_b}{3}+\\\frac{\eta_a\eta_b}{9}\Big[\text{cos}^2(\Phi)-6\Big]\Bigg)+(1-\eta_a)(1-\eta_b)P_{\text{dark}}^{2} \Bigg].
\end{multline}
For all the eight useful outcomes shown in Table \ref{sifting}, the expression for yield is found to be the same, and therefore, the overall yield is
\begin{align}\label{eq43}
    Y_{11}=&8Y_{11}^{(c,c,0)}.
\end{align}
Similarly, we can derive the QBER. We consider that the main cause of error is dark counts and calculate the QBER as,
\begin{multline}\label{singleqber}
    e_{11}Y_{11}=8(1-P_{\text{\text{dark}}})^4\Bigg[\frac{\eta_d\eta_a\eta_b}{18}\left(1+\text{cos}^2(\Phi)\right)+P_{\text{dark}}\Bigg(\frac{\eta_a+\eta_b}{3}\\+\frac{\eta_a\eta_b}{9}\Big[\text{cos}^2(\Phi)-6\Big]\Bigg)+(1-\eta_a)(1-\eta_b)P_{\text{dark}}^{2}\Bigg],
\end{multline}
For DPS-MDI-QKD with WCSs, the single-photon gain $Q_{11}$ is defined as the probability of observing a successful detection, given both Alice and Bob send single photons. Since the photon number distribution is Poissonian in a coherent state, the single-photon gain is given by,
\begin{align}\label{singlegain}
    Q_{11}=\mu_a\mu_be^{-\mu_a-\mu_b}Y_{11}.
\end{align}
Using the quantities derived in Eqs. (\ref{overgain}), (\ref{overqber}), (\ref{singleqber}) and (\ref{singlegain}), we derive the secure key rate \cite{ranu2021differential} as,
\begin{align}
    R_{\text{sec}}\geq~~\frac{1}{N}Q_{11}(1-H(e_{11}))-Q_m\times f\times H(E_m).
\end{align}
 The key rate plots using simulation parameters for DPS-MDI-QKD, same as used by \cite{ranu2021differential}, are shown in Fig. \ref{pol_mism}. It is found that the secure key rate reduces with increasing polarization mismatch. This happens due to the effect of polarization mismatch on the interference of the pulses arriving at the measurement setup. The secure key rate of \cite{ranu2021differential} is achieved for zero mismatches in polarization. No positive, secure key rate is found for a polarization mismatch of more than $11^{\circ}$.
\section{Effects of pulse-width mismatch on MDI protocols}\label{pulse_mismatch}
 Channel asymmetry between Alice and Charlie, and Bob and Charlie is the reasons for a pulse-width mismatch in MDI-QKD protocols. Along with the non-identical components such as WCSs, DLIs, and modulators used by Alice and Bob in an MDI setup, statistical fluctuations during operations can also result in a pulse-width mismatch. In this section, we quantify the effects of pulse-width mismatch on the performance of MDI-QKD, and re-emphasize the advantages of a plug-and-play architecture for MDI-QKD over a conventional MDI topology \cite{choi2016plug}. 

\subsection[Action of beamsplitter on pulses with different FWHM]{Action of beamsplitter on pulses with different full width at half maximum (FWHM)}\label{sec:BS action}
\begin{figure}%[ht!]
    \centerline{\includegraphics[width=0.6\textwidth,height=\textheight,keepaspectratio]{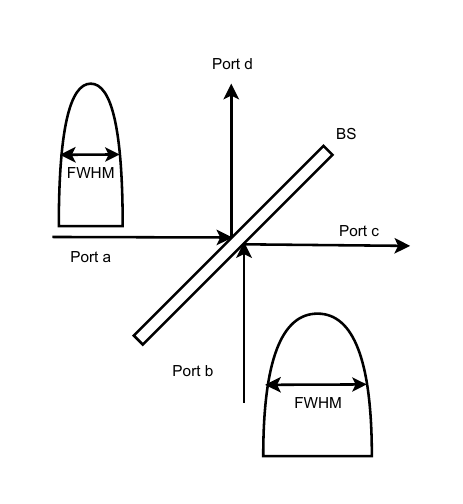}}
    \caption[Pulse-width mismatch at beamsplitter ports.]{Port $a$ and $b$ are the input ports, and port $c$ and $d$ are the output ports of the beamsplitter. Input pulses have different FWHM.}
    \label{fig:bs} 
\end{figure}

Fig.~\ref{fig:bs} shows two weak coherent pulses with different FWHM incidents on the input ports ($a$ and $b$) of a 50:50 beamsplitter. 
We want to obtain the quantum state at the beamsplitter output ports ($c$ and $d$). The states at the input ports of the beamsplitter can be written as
\begin{equation}
    \ket{\psi}_a=\text{exp}\left(\int_{-\infty}^{\infty}\left[\mu(t)\hat{a}^\dagger(t)-\mu^*(t)\hat{a}(t)\;\text{dt}\right]\right)\ket{0},\label{eq:port a}
\end{equation}
\begin{equation}
    \ket{\psi}_b=\text{exp}\left(\int_{-\infty}^{\infty}\left[\nu(t)\hat{b}^\dagger(t)-\nu^*(t)\hat{b}(t)\;\text{dt}\right]\right)\ket{0}. \label{eq:port_b}
\end{equation}
We define $\mu(t)$ and $\nu(t)$ such that $\vert \mu(t)\vert^2 $ and $\vert \nu(t)\vert^2 $ form a Gaussian distribution in time, with the same mean but different variance. Also, the definition of $\mu(t)$ and $\nu(t)$ should satisfy the following constraints
\begin{equation}
    \int_{-\infty}^{\infty}\vert \mu(t)\vert^2\;\text{dt}=\mu, 
\end{equation}
\begin{equation}
    \int_{-\infty}^{\infty}\vert \nu(t)\vert^2\;\text{dt}=\nu, 
\end{equation}
where $\mu$ and $\nu$ are the mean photon number of Alice and Bob's pulse, respectively. Hence, we write \cite{navarrete2018characterizing,zhang2020generalized}
\begin{equation}
    \mu(t)=\frac{\sqrt{\mu}}{(2\pi\sigma_a^2)^{\frac{1}{4}}}\text{exp}\left(\frac{-t^2}{4\sigma_a^2}\right)\text{exp}(i\phi_a-i\omega_a t),
\end{equation}

\begin{equation}
\nu(t)=\frac{\sqrt{\nu}}{(2\pi\sigma_b^2)^{\frac{1}{4}}}\text{exp}\left(\frac{-t^2}{4\sigma_b^2}\right)\text{exp}(i\phi_b-i\omega_b t),
\end{equation}
where $\phi_a$ and $\phi_b$ are the random phases of Alice and Bob's pulse, respectively, and $\sigma_a$ and $\sigma_b$ are their respective pulse widths. Without loss of generality, we integrate over $t\in(-\infty, \infty$) to obtain all the results in this section. We define the FWHM of the Gaussian pulse from its standard deviation by using the relation $\text{FWHM}=2\sigma\sqrt{2~\text{ln}2}$. When both $\sigma_a$ and $\sigma_b$ are equal, the two pulses have the same FWHM. In other words, the difference between $\sigma_a$ and $\sigma_b$ quantifies the extent of the pulse-width mismatch. The decoy-state method typically applies a random phase over the encoded pulses so as to prevent a photon-number splitting attacks \cite{lo2005decoy,lo2012measurement}. We also assume that the two wavepackets have different central frequencies, denoted by $\omega_a$ and $\omega_b$. We obtain the output state at the beamsplitter by applying the transforms shown in Eq. \eqref{eq17} to the input states shown in Eq. \eqref{eq:port a} and Eq. \eqref{eq:port_b} as
\begin{align}
\ket{\psi}_c=\text{exp}\Bigg(\int_{-\infty}^{\infty}\bigg[\frac{\mu(t)+\nu(t)}{\sqrt{2}}\hat{c}^\dagger(t)-\frac{\mu^*(t)+\nu^*(t)}{\sqrt{2}}\hat{c}(t)\;\text{dt}\bigg]\Bigg)\ket{0},\label{eq:port c}
\end{align}
\begin{align}
\ket{\psi}_d=\text{exp}\Bigg(\int_{-\infty}^{\infty}\bigg[\frac{\mu(t)-\nu(t)}{\sqrt{2}}\hat{c}^\dagger(t)-\frac{\mu^*(t)-\nu^*(t)}{\sqrt{2}}\hat{c}(t)\;\text{dt}\bigg]\Bigg)\ket{0}.\label{eq:port d}
\end{align}
Assuming $\mu=\nu$, we obtain mean photon numbers at the output ports of the beamsplitter as
\begin{equation}
\mu_c=\mu+\mu \sqrt{\frac{2\sigma_a \sigma_b}{\sigma_a^2+\sigma_b^2}}\; \text{exp}\left(\frac{\Delta\omega^2 \sigma_a^2\sigma_b^2}{\sigma_a^2+ \sigma_b^2}\right)\text{cos}~(\Delta\phi),
\end{equation}\label{eq:mean_c}
\begin{equation}
\mu_d=\mu-\mu \sqrt{\frac{2\sigma_a \sigma_b}{\sigma_a^2+\sigma_b^2}}\; \text{exp}\left(\frac{\Delta\omega^2 \sigma_a^2\sigma_b^2}{\sigma_a^2+ \sigma_b^2}\right)\text{cos}~(\Delta\phi),
\end{equation}\label{eq:mean_d}
where $\Delta \phi=\phi_a-\phi_b$ and $\Delta\omega=\omega_a-\omega_b$.

\subsection{Effect of pulse-width mismatch on HOM interference}\label{sec: HOM visibilty}
The visibility ($V_{\text{HOM}}$) of HOM interference is defined as
\begin{equation}
	V_{\text{HOM}}=1-\frac{p_{\text{cd}}}{p_c\times p_d},
\end{equation}\label{eq:visibility_def}
where $p_{\text{cd}}$ is the probability of coincidence counts and $p_c$ ($p_d$) is the probability of count at the output port $c$ ($d$) of the beamsplitter \cite{wang2017realistic}. We write $p_c$ as
\begin{align}
	p_c =& 1-\text{e}^{-\mu_c} \nonumber \\
	=& 1-\text{exp}\Bigg(-\mu-\mu \sqrt{\frac{2\sigma_a \sigma_b}{\sigma_a^2+\sigma_b^2}}\; \text{exp}\left(\frac{\Delta\omega^2 \sigma_a^2\sigma_b^2}{\sigma_a^2+ \sigma_b^2}\right)\text{cos}\Delta\phi\Bigg).
\end{align}
Averaging over uniformly distributed $\Delta\phi$, we obtain $p_c$ as
\begin{equation}
    p_c = 1-e^{-\mu}I_0\left(\mu\sqrt{\frac{2\sigma_a \sigma_b}{\sigma_a^2+\sigma_b^2}}\;\text{exp}\left(\frac{\Delta\omega^2 \sigma_a^2\sigma_b^2}{\sigma_a^2+ \sigma_b^2}\right)\right).
\end{equation}
Similarly, we obtain $p_d$ as
\begin{equation}
p_d= 1-e^{-\mu}I_0\left(-\mu\sqrt{\frac{2\sigma_a \sigma_b}{\sigma_a^2+\sigma_b^2}}\;\text{exp}\left(\frac{\Delta\omega^2 \sigma_a^2\sigma_b^2}{\sigma_a^2+ \sigma_b^2}\right)\right),
\end{equation}
%\vspace{0.5cm}
and get the coincidence probability as
\begin{align}
p_{\text{cd}}=& 1-2e^{-\mu}I_0\left(-\mu\sqrt{\frac{2\sigma_a \sigma_b}{\sigma_a^2+\sigma_b^2}}\;\text{exp}\left(\frac{\Delta\omega^2 \sigma_a^2\sigma_b^2}{\sigma_a^2+ \sigma_b^2}\right)\right)+e^{-2\mu}.
\end{align}
 Here, $I_{0}(x)$ is the modified Bessel function of the first kind.
 \begin{figure}%[h!]
    \centerline{\includegraphics[width=0.6\textwidth,height=\textheight,keepaspectratio]{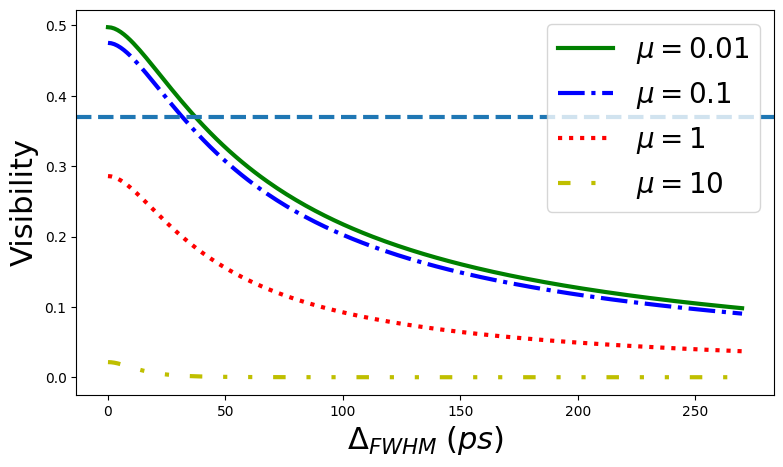}}
    \caption[Variation in the visibility of HOM interference with FWHM.]{Variation in the visibility of HOM interference with FWHM for different mean photon numbers.}
    \label{fig:vis_fwhm} 
\end{figure}
 We set $\Delta\omega=0$ and plot the visibility of HOM interference versus the difference in FWHM of the input pulses. From Fig.~\ref{fig:vis_fwhm}, we observe that the visibility of HOM interference decreases as the FWHM mismatch between the pulses increases. As $\mu$ decreases, multi-photon emission probability decreases, thereby approaching the maximum theoretical visibility of 0.5 for a very low mean photon number of $0.01$ \cite{yuan2014interference}.  
			
We have represented the threshold visibility ($V_{\text{th}}$) of $0.37$ using a dashed horizontal line in Fig.~\ref{fig:vis_fwhm}. Typical QKD implementations use a mean photon number of $0.1$. As expected, a higher mean photon number yields a HOM visibility below $V_{\text{th}}$, thereby leading to $0$ secure key rate for the polarization-based MDI-QKD. We observe that $V_{\text{HOM}}\to V_{\text{th}}$ when $\Delta_{\text{FWHM}}$ is around $30$ ps for $\mu=0.1$. Let $R_{\text{max}}$ be the secure key rate of polarization-based MDI-QKD under a perfect HOM interference using two independent WCS. It has been shown that the secure key rate of the polarization-based MDI-QKD drops down to $50\%$ of $R_{\text{max}}$ when the $V_{\text{HOM}}$ reduces from $0.5$ to $0.45$ and reduces down to $10\%$ of $R_{\text{max}}$ for $V_{\text{HOM}}=0.4$ \cite{yuan2014interference}. We observe from Fig.~\ref{fig:vis_fwhm} that $\Delta_{\text{FWHM}}$ of $12$ ps and $24$ ps leads to $V_{\text{HOM}}=0.45$ and $V_{\text{HOM}}=0.4$, respectively.   

Channel length asymmetry between Alice and Charlie, and Bob and Charlie is one of the reasons behind the pulse-width mismatch. We have estimated the degree of asymmetry in fiber length that would result in poor HOM visibility. If we assume that group velocity dispersion is the primary reason behind pulse broadening, and the pulses from Alice and Bob encounter different channel lengths, we obtain the pulse-width mismatch
\begin{equation}
    \Delta_{\text{FWHM}} = D (\Delta L) (\Delta\lambda),
\end{equation}
where $D$ is the dispersion parameter, $\Delta L$ is the asymmetry in the fiber channel length, and $\Delta\lambda$ is the spectral width of the source \cite{agrawal2012fiber}. Typically, $D$ is around $17$~ps/nm-km for optical fibers at telecommunication wavelength, and $\Delta\lambda$ can be around $1$pm-$10$ pm for the commercially available lasers that are used in QKD implementations. In our calculations, we assume $\Delta\lambda$ equals $10$ pm. In Table~\ref{table2}, we tabulate the extent of channel asymmetry needed to observe a drop in the secure key rate. Our calculations show the non-trivial effect of channel length asymmetry on the secure key rates of polarization-based MDI and the need for minimizing any such asymmetry in an experimental MDI implementation. Fortunately, the effect of channel length asymmetry can be easily mitigated by adding extra fiber to the shorter channel or by using suitable lengths of dispersion compensating fibers, and thereby providing a secure key rate identical to that of an ideal MDI setup.
\begin{table}[h]
\caption[HOM interference visibility for different channel length asymmetry.]{Estimates for length and pulse-width mismatch corresponding to typical values of $\text{V}_{\text{HOM}}$ and secure key rate.}\label{table2}%
\begin{tabular}{@{}cccc@{}}
\toprule
\textbf{$\text{V}_{\text{HOM}}$} & \textbf{Secure key rate} & \textbf{$\Delta_{\text{FWHM}}$ (ps)} & \textbf{$\Delta L$ (km)} \\
\midrule
$0.5$ & $R_{\text{max}}$ & $0$ & $0$ \\
\hline
$0.45$ & $0.5 R_{\text{max}}$ & $12$ & $70.6$  \\
\hline
$0.4$ & $0.1 R_{\text{max}}$ & $24$ & $141.2$ \\
\hline
$0.37$ & $0$ & $30$ & $176.5$ \\
\botrule
\end{tabular}
\end{table}
\section{Plug-and-play DPS-MDI-QKD protocol} \label{Pnp_protocol}
\begin{figure}[ht!]
    \centering
    \includegraphics[width=0.5\linewidth]{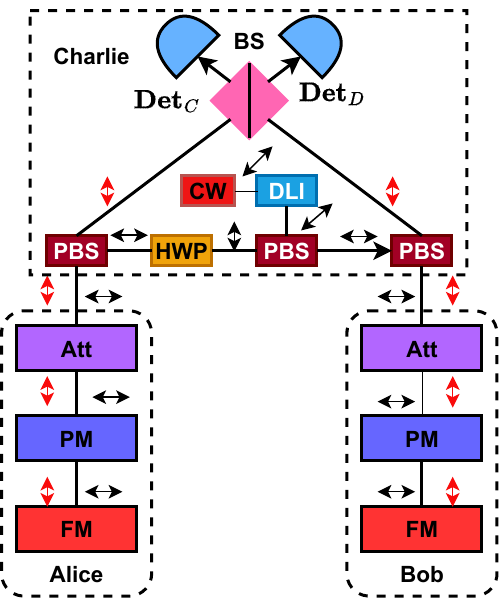}
    \caption{Schematic of plug-and-play DPS-MDI-QKD. CW: continuous wave laser, DLI: delay line interferometer, PBS: polarization beamsplitter, HWP: half-wave plate, BS: beamsplitter, 
    Det: single-photon detector, PM: phase modulator, Att: attenuator, FM: Faraday mirror. The polarization of modes to and from Alice (Bob) are shown in black and red arrows, respectively. A random number generator is supposed to be there within the PM block.}
    \label{schematic}
\end{figure}

In this section, we present the plug-and-play version of the DPS-MDI-QKD protocol, also patented as \cite{patent}. As shown in Fig. \ref{schematic}, Alice and Bob are assumed to be two legitimate parties willing to establish a secure key. Charlie is the third party with a coherent laser source and measurement setup, and can also be a potential eavesdropper. We consider that Alice and Bob are at an equal distance from Charlie (symmetric configuration). The protocol works as follows.
\begin{enumerate}
    \item Charlie generates 45$^{\circ}$ polarized coherent pulses in three-time bins using laser.
    \item He splits them into two identically polarized pulses using a polarizing beamsplitter (PBS) and a half wave plate (HWP) on one output arm of PBS.
    \item Charlie transmits these two identical pulses, one to Alice and the other to Bob.
    \item Alice (Bob) keeps her (his) phase modulator (PM) and attenuator (Att) off and lets the pulse from Charlie reflect back from the Faraday mirror (FM).
    \item Alice (Bob) chooses a random phase value in $(0,2\pi)$.
    \item Alice (Bob) generates two random bits using a random number generator.
    \item Alice (Bob) randomizes the phase of the reflected pulse in three time bins from FM and encodes the generated bits as phase differences between consecutive time bins using PM. He (she) keeps record of the two bits encoded.
    \item Alice (Bob) attenuates the pulses to a single-photon level using an attenuator and transmits it back to Charlie.
    \item Charlie performs a Bell state measurement using a beamsplitter and detectors as shown in Fig. \ref{schematic}.
    \item As discussed in the sifting scheme in Table \ref{sifting}, Charlie announces only the useful outcomes. Other outcomes are discarded. 
    \item For all the announced outcomes, Alice and Bob announce their random phase values. The sifting is done only if their random phase values are the same. 
    \item For all these results, Alice and Bob sift the key bits as shown in Table \ref{sifting}.
    \item Bob reveals his sifted key bits at some random instants announced by Alice. Alice finds quantum bit error rate (QBER) by comparing her corresponding bits with them.
    \item If QBER is less than a predefined threshold based on the inherent error rate, they perform error reconciliation and privacy amplification on the remaining sifted key bits. They discard the bits announced during QBER estimation.
    \item Otherwise, they discard the round and start a new one after some time.
\end{enumerate}
\subsection{Robustness of plug-and-play DPS MDI}
We have discussed the effects of channel asymmetry on DPS-MDI-QKD in Section \ref{pol_mismatch} and Section \ref{pulse_mismatch}. Plug-and-play DPS-MDI-QKD resolves such issues. Electric field polarization in single-mode fibers is governed by the birefringence of the fiber material. Environmental fluctuations and resultant stress on the fiber are slowly varying phenomena that affect the secure key rate \cite{muller1996quantum}. The Faraday mirror transforms the polarization of the input state to an orthogonal polarization and reflects it through the same fiber. Therefore, the signal traveling forward from Charlie to Alice (Bob) and reflected back will undergo an adjustment, resulting in better visibility of interference compared to DPS-MDI-QKD, and compensates for any birefringence effects \cite{muller1997plug}. Therefore, our plug-and-play scheme improves the performance of DPS-MDI-QKD in the presence of polarization mismatch, as both the signals are generated from the same laser and sent to Alice and Bob. While returning to Charlie through the same fiber, they undergo self-alignment of polarization, thus mitigating the effect of polarization mismatch.\par
OPLLs are required to share a phase reference in the conventional MDI implementations where Alice and Bob use independent lasers. Further, Alice and Bob use independent delay lines to create three-time bins in the DPS-MDI protocol. The time-bins of Alice and Bob may not be identical due to differences in the design of their respective DLIs, leading to pulse-width mismatch. The use of a single DLI and laser in the proposed plug-and-play architecture removes the requirement of an OPLL and eliminates the possibility of pulse-width mismatch arising from statistical fluctuations in Alice and Bob's devices. Therefore, the plug-and-play architecture leads to a simpler cost-effective setup by reducing the number of components required for implementation. As explained in the previous sections, there exist many factors resulting in channel asymmetry in MDI-QKD protocols. Plug-and-play architecture takes care of many of the causes of channel asymmetry and, hence, can achieve the secure key rate of the ideal MDI-QKD. However, even in the plug-and-play architecture, special measures must be taken to compensate for the asymmetric channel length. Channel length asymmetry can be removed in a simple and cost-effective way by adding an extra fiber length to the shorter channel. Alternatively, one can use negative dispersion fiber of appropriate length to cancel the pulse broadening arising due to the positive dispersion coefficient of the commercial fibers typically used in MDI implementations. In the scenario when all such measures to remove channel asymmetry have been implemented, the proposed plug-and-play architecture achieves the secure key rate identical to that of an ideal DPS-MDI-QKD.
%\section{Result}\label{result}

\section{Conclusion}\label{conclusion}
In this work, we have analyzed the practical security of the recently proposed DPS-MDI-QKD protocol. The implementation of any QKD scheme is affected by various channel impairments. Here, we have quantified the effect of polarization mismatch on the performance of the DPS-MDI-QKD protocol, and found that it can withstand a polarization mismatch of at most 11$^\circ$. For a polarization mismatch of more than 11$^\circ$, no positive secure key rate is found. On the other hand, a non-similar pulse broadening in an asymmetric configuration of  MDI-QKD only marginally affects the HOM visibility. In Table \ref{table2}, we see that a channel length asymmetry of $176$ km leads to zero secure key rate for a polarization-based MDI protocol.  Using a common laser source and DLI reduces the chances of various misalignments, such as polarization and pulse-width mismatch, leading naturally to our proposed plug-and-play DPS-MDI-QKD protocol. We note that the pulse width mismatch arising only due to non-identical components is taken care of by the presented plug-and-play architecture.\par 
We have also presented a modified sifting scheme for DPS-MDI-QKD (applicable to the plug-and-play version), which improves the sifting rate from $\frac{4}{9}$ to $\frac{2}{3}$ and can be readily used with the presented plug-and-play version. This provides an improvement in the secure key rate of the scheme.\par
Going forward, it is important to quantify the effect of pulse-width mismatch for other MDI protocols, such as DPS-MDI-QKD and twin-field QKD. Other directions for future research are to prove the full security of this scheme with an unreliable source \cite{xu2015measurement, zhao2008quantum} in asymptotic and finite size key regimes, and finding the effect of finite decoy states on the DPS-MDI-QKD scheme.\par
Recently, a quantum secure direct communication (QSDC) protocol using an entangled pair and single-photon sources in MDI architecture has been presented in \cite{niu2018measurement}.  It would be interesting to explore the possibility of getting QSDC protocol using our plug-and-play DPS-MDI-QKD. This will reduce the cost of implementation of the QSDC scheme, but security consequences need to be analyzed in a practical scenario. We have also recognized during this work that there is a similarity between Twin-Field (TF) QKD and MDI-QKD \cite{sharma2024twin}. The difference comes from the need for single photon interference to sift keys in TF-QKD. Therefore, we will explore the possibility of extending DPS-MDI-QKD to the corresponding TF-QKD version.
\bmhead{Acknowledgements} 
We are grateful to the Mphasis F1 Foundation, and the Ministry of Electronics and Information Technology (MeitY), Government of India and Bharat Electronics Limited (BEL), India for the financial support to this work. NS would also like to thank the I-hub Quantum Technology Foundation (QTF), Department of Science and Technology (DST), India, for the Chanakya Fellowship.
\bibliography{main}
\end{document}